# Intracluster light is already abundant at redshift beyond unity

Hyungjin Joo[1] & M. James Jee[1,2]
[1]Department of Astronomy, Yonsei University, Seoul, Republic of Korea.
[2]Department of Physics and Astronomy, University of California, Davis, Davis, CA, USA

Intracluster light (ICL) is diffuse light from stars that are gravitationally bound not to individual member galaxies, but to the halo of galaxy clusters. Leading theories[1,2] predict that the ICL fraction, defined by the ratio of the ICL to the total light, rapidly decreases with increasing redshift, to the level of a few per cent at $z > 1$. However, observational studies have remained inconclusive about the fraction beyond redshift unity because, to date, only two clusters in this redshift regime have been investigated. One shows a much lower fraction than the mean value at low redshift[3], whereas the other possesses a fraction similar to the low-redshift value[4]. Here we report an ICL study of ten galaxy clusters at $1 \lesssim z \lesssim 2$ based on deep infrared imaging data. Contrary to the leading theories, our study finds that ICL is already abundant at $z \gtrsim 1$, with a mean ICL fraction of approximately 17%. Moreover, no significant correlation between cluster mass and ICL fraction or between ICL colour and cluster-centric radius is observed. Our findings suggest that gradual stripping can no longer be the dominant mechanism of ICL formation. Instead, our study supports the scenario wherein the dominant ICL production occurs in tandem with the formation and growth of the brightest cluster galaxies and/or through the accretion of preprocessed stray stars.

## Main

Intracluster light (ICL) is predominantly distributed in the central region of the cluster, in most cases around the brightest cluster galaxy (BCG) out to several hundred kilo-parsecs[3-6]. Some studies reported that significant ICL is also found around intermediate and massive satellites[7,8]. We detected ICL around the BCGs of ten galaxy clusters at $z \gtrsim 1$with the Wide Field Camera 3 (WFC3) near-infrared imager on board Hubble Space Telescope (HST) (Fig. 1). In most cases, a clear surface brightness (SB) profile is obtained out to approximately 200 kpc, where it approaches the surface brightness limit $\mu \approx 28$ mag arcses$^{-2}$ (Fig. 2). The exception is the result for JKCS041, which is the highest redshift ($z = 1.8$) target in our sample (Extended Data Table 1). Its SB profile approached the limit at around 100 kpc. Figure 2 shows that, overall, the SB profiles are well described by a superposition of two or three multi-Sérsic components convolved with the instrument point spread function (PSF). Regardless of the number of components, the outermost component is predominantly responsible for the shape of the outer part of the SB profile, which is assumed to characterize the ICL here, whereas all inner components (one component if the total number of components is two) are considered to represent the BCG profile. Most clusters in our sample show no significant gradients in their SB colour profiles and their colours are in good agreement with those of the reddest cluster members. Exceptions are found for SpARCS1049 and IDCS1426, which possess a clear negative gradient, with the colour difference between the BCG- and ICL-dominant regions being around 1 mag (Fig. 3). IDCS1426 and SpARCS1049 are the second ($z$ = 1.75) and third ($z$ = 1.71) highest redshift clusters in our sample. When converted to the rest-frame $B$ and $V$ mags, the BCG colours span the $0.5 < B - V < 0.8$ range, which overlaps the theoretical distribution[9]. Combining spectroscopic and photometric member selections, we measured the BCG + ICL and ICL fractions ($f_{\mathrm{BCG+ICL}}$ and $f_{\mathrm{ICL}}$) using an aperture of $r$ = 0.5 Mpc (Extended Data Table 2). The mean BCG fraction is approximately 4.5%, which is well bracketed by the values in previous studies[3,8,10,11]. Figure 4 shows that the mean ICL fraction of our sample is similar to that of the low-redshift sample in the literature[1,3,5,8,10-21].

One potential difficulty for the interpretation of Fig. 4 is the diversity of the methodology in the previous studies. We investigated the impacts of the following two factors: aperture size and ICL definition. The results compiled in Fig. 4 are based on apertures ranging from 100 kpc to 1.7 Mpc. We verified that there is no correlation between aperture size and ICL fraction in the published result. Moreover, the mean aperture size in the literature is 0.58 Mpc, which is similar to our choice of 0.5 Mpc. Finally, when we repeated the analysis using the subsample that used the aperture sizes between 0.35 and 0.65 Mpc, the result remained unchanged. The ICL community is aware that the results from the traditional SB cut (SBC) method can differ systematically from those obtained by the new multicomponent decomposition method[22] to which our approach belongs. To address the issue, we divided the literature sample into the SBC and multicomponent decomposition subsamples. Our regression based on the latter shows that the slope is still consistent with zero at the $2\sigma$ level.

We find that dwarf galaxies fainter than our detection limit do not bias our ICL fractions high. To investigate the impact of sources fainter than our detection limit, we carried out image simulations by randomly distributing dwarf galaxies, whose number is estimated by fitting a Schechter luminosity function to the detected source distribution and computing the difference between the best-fit luminosity function and observed distribution. We considered two types of radial distributions. The first is a uniform distribution across the field. The second is the distribution that follows a Navarro–Frenk–White profile[23]. In the first case, the ICL fraction is unchanged because adding a uniform dwarf galaxy distribution is equivalent to elevating the sky level by the same degree simultaneously across the entire field. In the second case, the dwarf galaxies are mostly concentrated near the BCG. Although this certainly would lead to the overestimation of the BCG luminosity, the impact on the ICL luminosity was negligible.

We rule out the possibility that unmasked galaxy light might artificially increase the ICL fraction. In our analysis, we employed a moderate-sized mask and later applied a correction factor to obtain the result effectively measured with the full mask (see Methods). This correction scheme was verified to be accurate, leading to only an approximately 0.02% difference in the average ICL fraction (Extended Data Table 2).

One may argue that the ten clusters in our sample correspond to the most massive population at high redshift and thus should not be compared directly with the low-redshift clusters. Although four of our ten clusters may potentially belong to extremely massive (around $10^{15} M_\odot$) populations in the $1 \lesssim z \lesssim 2$ universe, the masses of the remaining six clusters span the range 2–6 × $10^{14} M_\odot$ (Extended Data Table 3). We found that the cluster masses do not correlate with the ICL fractions for our sample (Extended Data Fig. 5). Although theoretical studies[1,2,24–27] remain inconclusive about the $f_{ICL}$–mass correlation, observational studies[22,28,29] agree that there is no correlation. This lack of the $f_{ICL}$–mass correlation is also supported in the study where the sample is limited to a narrow redshift range of $z < 0.07$[22]. Hence, we do not attribute the absence of the $f_{ICL}$–redshift correlation to a selection effect.

The dominant ICL production mechanism is still unknown, although the current consensus is that merger, stripping and preprocessing are the three important candidates[22,30–32]. The ICL fraction is an important observable sensitive to the timescale of the ICL formation, and its evolution with redshift can be used to discriminate between competing theories regarding the dominant ICL production mechanism. A strong evolution[1,2,10,28,33] favours a gradual process through stripping, whereas the opposite[5,29] supports the scenario wherein the dominant ICL production happened at high redshifts. The absence of the apparent evolution of the ICL fraction in the $0 \lesssim z \lesssim 2$ redshift regime in the current study contracts the current leading theories[1,2], which predict that the mean ICL fraction decreases to a negligible level (less than 5%) at $z = 1.5$ (Fig. 4). Therefore, the most straightforward interpretation of the current finding is that the dominant ICL formation and its evolution with redshift can occur not through gradual stripping, but in tandem with the BCG formation and growth, and/or through the accretion of preprocessed stray stars.

Together with the ICL fraction, another traditional but still critical method to discriminate between competing theories on the dominant mechanism of ICL production is to investigate the ICL stellar population with its colour and compare the results with those of the cluster galaxies, including the BCG. For instance, if major mergers with the BCG are the dominant mechanism, no significant colour difference between BCG and ICL is expected. On the other hand, if the ICL is formed by a more gradual process such as stellar stripping, we expect that the ICL would be bluer than the BCG, or a

nega- tive gradient would be present in the radial colour profile. In this case, matching the colour between the ICL and cluster galaxies can constrain the progenitors of the ICL. Previous observational studies in general support the presence of negative gradients, although exceptions are not uncommon[4,19,34,35], which implies that the gradient may depend on the particular assembly history of individual clusters[9]. The absence of the ICL colour gradient in most of the cases in our sample indicates that gradual stripping is not likely to be the dominant mechanism of ICL production within the $1 \lesssim z \lesssim 2$ epoch. It is possible that occasional major mergers can potentially mix the intracluster stars and flatten the ICL colour profile even in the case where ICL production through stripping is dominant. However, as the ICL colours in the flat gradient cases are in good agreement with those of the reddest cluster members, our observation cannot reconcile with this scenario.

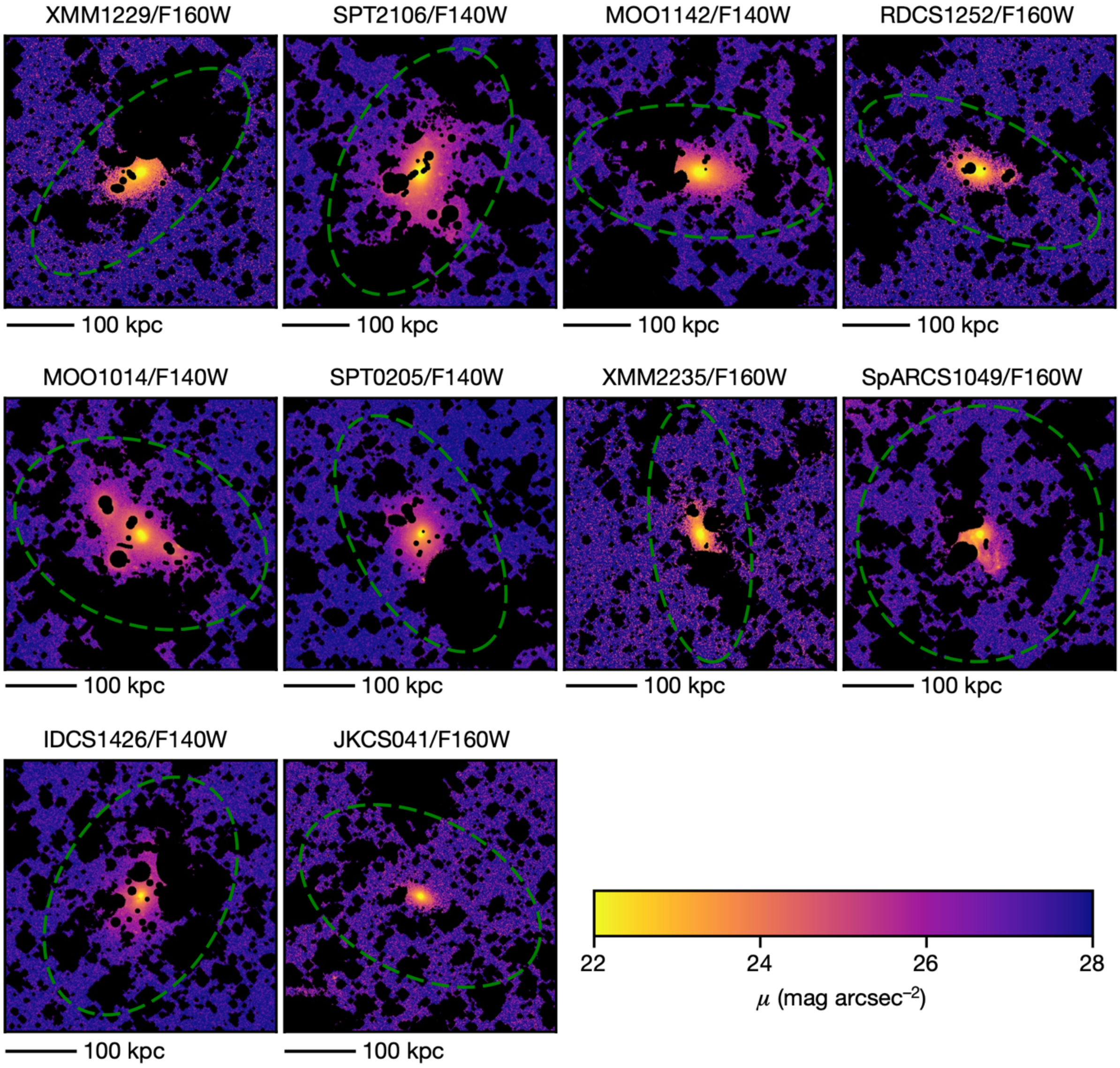

**Fig. 1 | BCG + ICL images of our ten $z \gtrsim 1$ clusters.** The images were created by masking out every discrete source detected by SExtractor except for the BCGs. Here the result is based on an expansion coefficient of 2. We also visually scanned the result and applied additional manual masking for the objects that SExtractor failed to identify. Green dashed lines show the SMA = 200 kpc ellipses, whose ellipticities and position angles are determined by AutoProf. Throughout the paper, we assume a flat Λ-dominated cold dark matter cosmology characterized by $h = 0.7$ and $\Omega_{m,0} = 1 - \Omega_{\Lambda,0} = 0.3$, where $h$, $\Omega_{m,0}$ and $\Omega_{\Lambda,0}$ represent the dimensionless Hubble, matter density, and dark energy density parameters at present day, respectively.

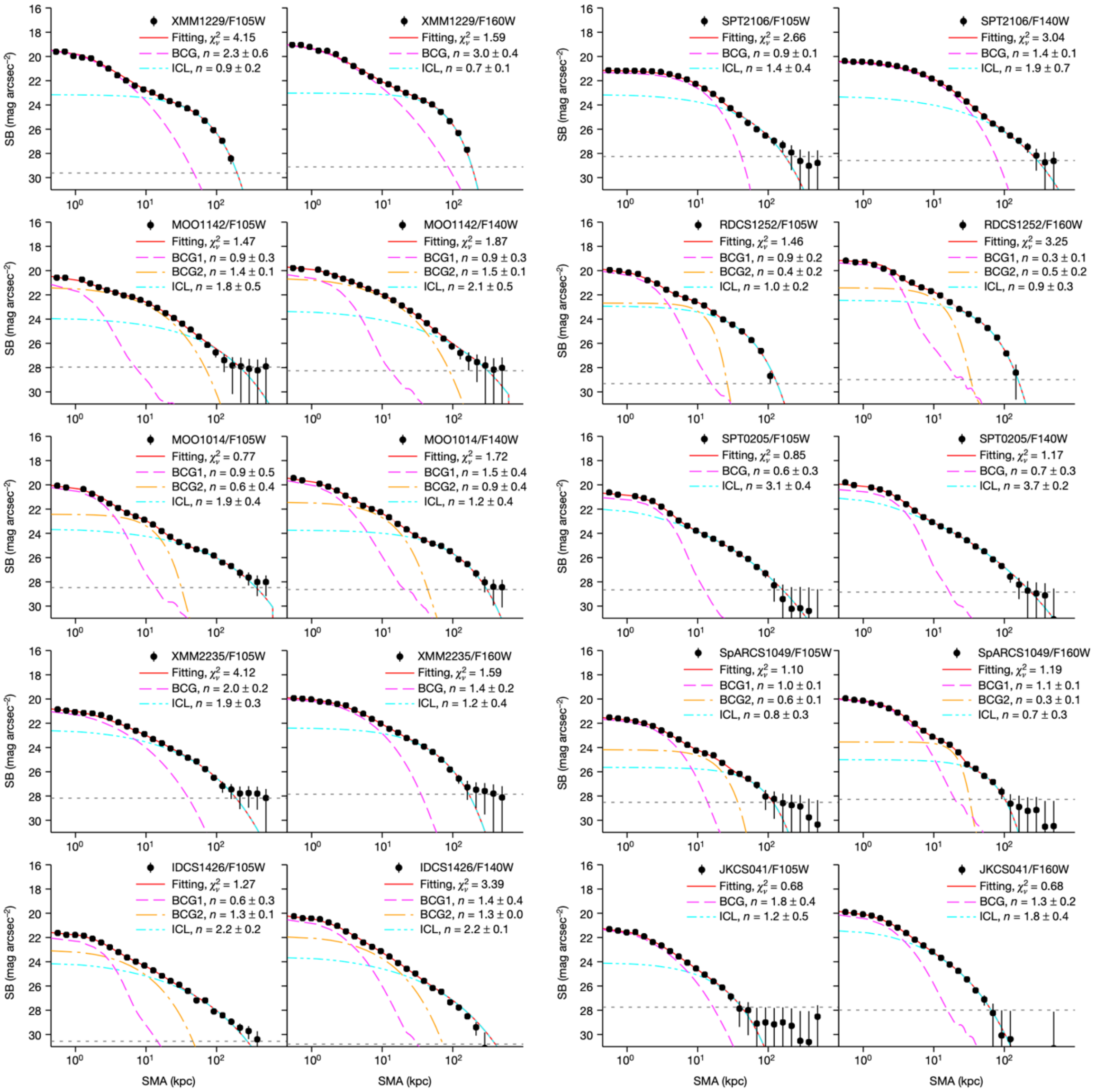

**Fig. 2 | BCG + ICL radial profiles.** Data points are the observed surface brightness from the elliptical bins. The errors are computed from quadratic sums of the background level error and shot noise. With dashed and solid lines, we show our best-fit multi-Sérsic component model. Magenta (cyan) dashed lines are the innermost (outermost) components. When the target requires three components, we use orange lines to represent the middle component. Red solid lines illustrate the summation of all components. The legends of each panel show the target name, filter type, $\chi^2$ value and best-fit Sérsic indices.

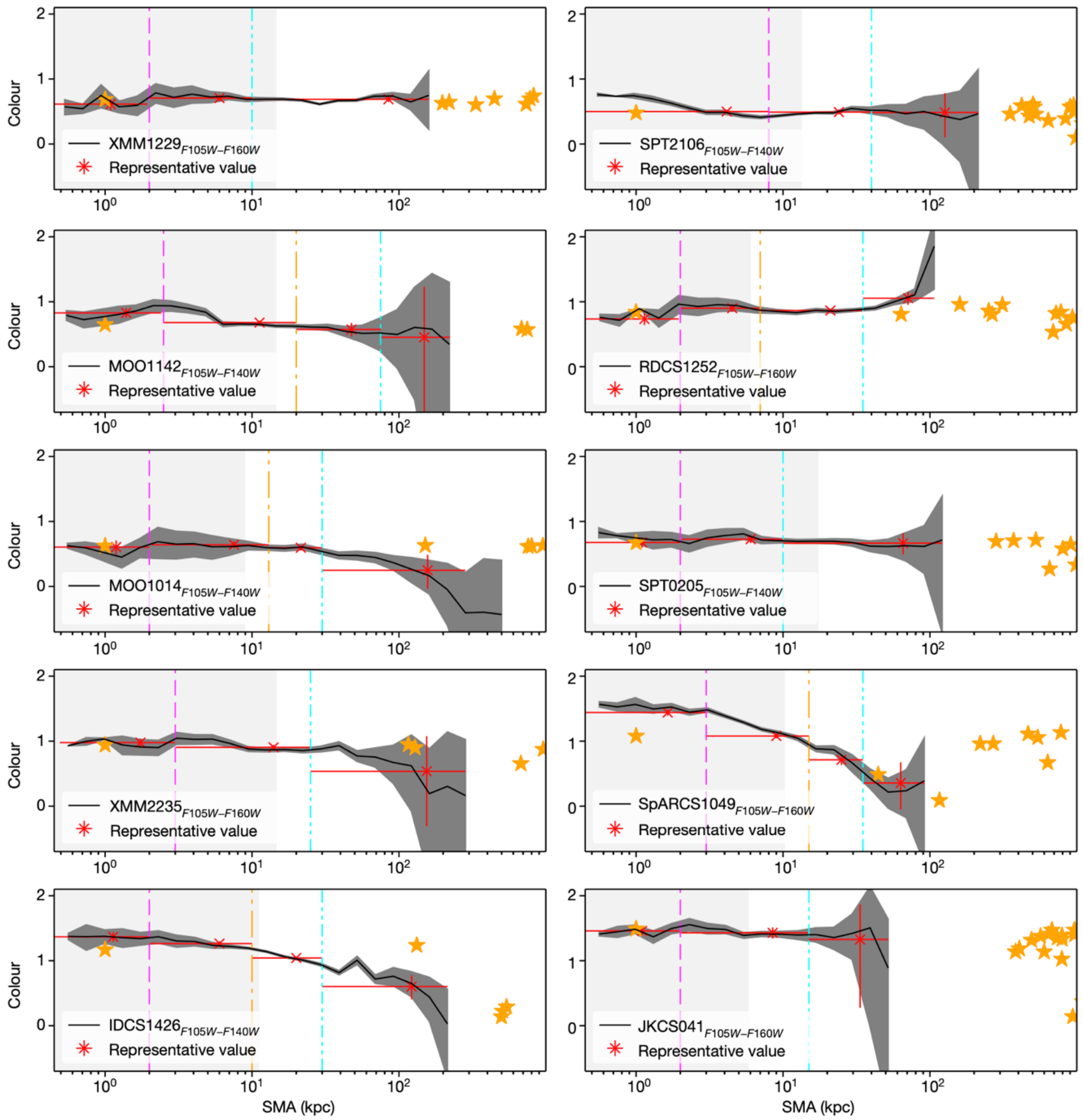

**Fig. 3 | BCG + ICL radial colour profiles.** Black solid lines are the observed colour. The dark grey shades represent the 68% uncertainty. The scale radius of each component is shown with the same colour scheme used in Fig. 2. The red data point is a representative mean value in each subregion. The orange stars indicate the colours and positions of spectroscopic member galaxies. The light grey shade indicates the radial extent of the BCG measured by SExtractor.

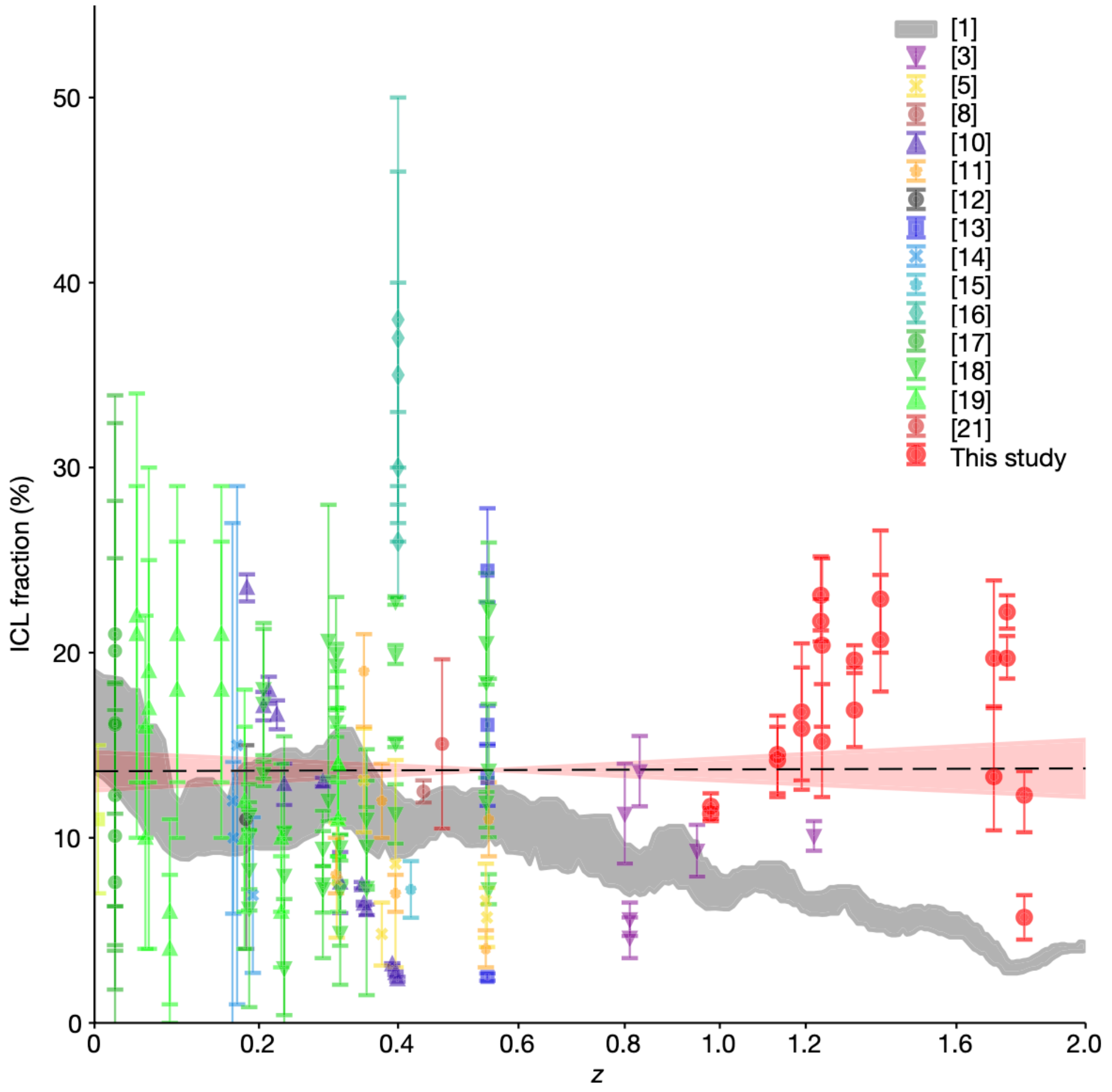

**Fig. 4 | ICL fraction evolution.** Filled red circles are the current results based on the $r$ = 0.5 Mpc aperture. We extrapolated the best-fit Sérsic profiles to the same aperture to estimate the total ICL flux. When we avoided the extrapolation and performed integration only within the range where the SB profile is above the detection limit, the resulting ICL fraction is reduced by about 1.9% on average. The mean aperture size of the literature sample is 0.58 Mpc. The dashed line and pink shade show the best-fit linear regression and its 68% uncertainty, respectively. We weighted all data points equally and adjusted them in such a way that the reduced $\chi^2$ value becomes unity. The comparison between our high-redshift and the literature low-redshift samples shows that there is no significant evolution of the ICL fraction with redshift in observation, which contradicts the current theoretical prediction[1] (grey). Although here we displayed the theoretical model that estimates the ICL fraction based on the SBC at 26.0 mag arcsec$^{-2}$, similarly steep evolutions are obtained even when different ICL definitions such as binding energy criteria are used.

# Methods

## Target selection

We searched the Mikulski Archive for Space Telescope for the WFC3-near-infrared imaging programs that have observed $z \gtrsim 1$ clusters in at least two filters with the surface brightness limit $\mu \approx 28$ mag arcsec$^{-2}$. We excluded the targets if they do not possess any distinct BCGs, or very bright stars/foreground galaxies are present near the cluster centres. The search resulted in a sample of ten galaxy clusters at $1 \lesssim z \lesssim 2$ with a minimum (maximum) redshift of 0.98 (1.803). Extended Data Table 1 summarizes our target selection, including the redshift, coordinate, program number, surface brightness limit and so on. Although F105W exists for all ten clusters, either F140W or F160W is available for each cluster.

## Reduction pipeline optimized for ICL measurement

The common data reduction procedure recommended in the HST Data Handbook[34] works well if one is interested in discrete astronomical sources such as stars and galaxies. However, when we are looking for signals from diffuse components whose surface brightness is within a subpercentage of the sky brightness and slowly varies across the detector, additional care is needed. Our reduction begins with the flat-fielded (FLT) images processed by the Space Telescope Science Institute calwf3 tool[36], which removes most instrumental signatures of WFC3, except for geometric distortion. We visually inspected these FLT images and manually masked out any remaining artefacts such as satellite/asteroid trails. We ran the TweakReg package[37] for astrometric calibration by finding common astronomical sources. The FLT images are already flatfielded with the default Space Telescope Science Institute composite flats, which are claimed to be accurate within less than 0.5%, except for the region within 128 pixels of the detector edge[34]. This claim has been verified by independently constructing residual flats utilizing large WFC3 survey programs[4]. We also investigated the impact of the residual flat in the final mosaic by performing drizzling as if we were stacking science frames and found that the dithered residual flats would cause at most around 0.4% errors, which is already negligible compared to other sources of errors (for example, background determination). In this study, we applied these residual flats to our FLT data to further reduce the residual flat errors. The application of the aforementioned residual flatfielding cannot remove large-scale sky gradients arising from intrinsic sky gradient, detector persistence, internal reflection and so on. We removed the sky gradient by fitting a first-order polynomial plane ($F(x,y) = ax + by + c$) to the object-masked residual-flatfielded image and subtracting the result. The subtraction result was visually inspected, and we discarded the frame if the first-order polynomial plane could not adequately describe the sky gradient. To create a final deep mosaic where ICL is measured, careful and consistent sky subtraction from each exposure is required. In typical ground-based data reduction designed for non-ICL-related studies, position-dependent sky estimation/subtraction is routinely performed after astronomical objects are masked out. Although this scheme may provide cosmetically ‘good’ results when different frames are combined together, the inevitable consequence is sky oversubtraction in the region where non-discrete astronomical components are dominant. Therefore, in this study, we determined only the global sky level (that is, a single constant) for each frame and subtracted it. Because individual exposures observed with dithers cover non-identical areas around the target, it is necessary to choose the same physical sky region that is present in common for all exposures. To accomplish this, we set up the largest possible annulus that is approximately centred on the BCG and is observed by all exposures. An illustration of this common annulus is presented in Extended Data Fig. 1.

Discrete astronomical sources in this annulus were detected using SExtractor[38] with the settings of DETECT_MINAREA = 5 and DETECT_ THRESHOLD = 2. As a matter of course, the resulting segmentation map fails to include the contributions from the faint diffuse wings of the objects. To address this, we gradually expanded the segmentation maps and investigated the resulting background level change as a function of the mask size. One should not increase the segmentation map by expanding the boundaries at the same rate for all objects. That is, the segmentation boundaries of compact objects should expand slower than those of extended objects. Thus, we used SExtractor’s half-light radius $r_h$ and scaled the expansion with it. The width ($w$) of the expansion band is determined by $w = c_e\, r_h$, where $c_e$ is the expansion coefficient. We stopped the segmentation map expansion at $c_e = 6$, beyond which the sky estimation converged (Extended Data Fig. 2). Note that this exposure-by-exposure sky level estimation is performed on the individual drizzled images to minimize the impact of the geo- metric distortion. We created the final mosaic using AstroDrizzle[33] with its sky subtraction option turned off. The output pixel scale is set to 0.05 arcsec. We used the Gaussian kernel for drizzling. Extended Data Fig. 3 schematically summarizes our data reduction pipeline.

## Object masking for ICL measurement

To characterize ICL from the mosaic image, it is necessary to mask out light from discrete objects. In the background level estimation above, we found that the choice $c_e = 6$ was sufficient. However, this large expansion coefficient cannot be used in the central region of the cluster because doing so would leave very few pixels there, resulting in too large statistical errors exceeding the systematic error caused by the incomplete masking (Extended Data Fig. 2). Therefore, for ICL measurement from the final stack, we chose to employ $c_e = 2$ for object masking and apply a correction factor to obtain the result effectively measured with the full mask ($c_e = 6$). The correction factor is derived by comparing multiple ICL profiles measured with different $c_e$ values. We masked out every discrete source detected by SExtractor except for the BCG. We also visually scanned the result and applied manual masking for the objects that SExtractor failed to identify (Fig. 1).

To assess the validity of our statistical correction scheme, we repeated our analysis with the full ($c_e$ = 6) masking. This is supposed to generate results with better accuracy at the expense of precision (that is, smaller systematic errors and larger statistical errors). However, if any large systematic discrepancy from our fiducial measurement (for example, statistically corrected result after the use of $c_e$ = 2) is found, this indicates that the aforementioned correction scheme is problematic. The mean of the differences is close to zero (about 0.02%, Extended Data Table 2), which verifies that our correction scheme with the use of the moderate masking expansion ($c_e$ = 2) is robust.

## Radial profile measurement with elliptical binning

As the BCG + ICL isophotes are elliptical, the use of circular binning would spread the BCG-ICL transition over multiple bins. Thus, we measured the radial profile of the BCG + ICL using an elliptical bin- ning scheme. To determine the ellipticity and position angle (PA) of the ellipse, we used the AutoProf package[39]. With both detection image and masking map as inputs, AutoProf calculates ellipticity and PA based on isophotal fitting and Fourier analysis.

Two sets of outputs are generated by AutoProf. One is a series of ellipticity and PA values, which vary with radius. The other is a single pair of ellipticity and PA, which represents the global shape. The former is useful when one's interest is the radius-dependent isophotal shape of high signal-to-noise-ratio objects. In this study, where our scientific interest is faint diffuse light, we use the second set of outputs.

We defined a series of semi-major axes (SMA) with a logarithmic scale and measured the surface brightness at each radial bin. We applied the $3\sigma$ clipping method to minimize the impact of the outliers and adopt the median as the representative surface brightness of the bin. This also reduces the effects of any potential unidentified substructures within each elliptical annulus. The total error of the surface brightness estimate is computed as the quadratic sum of the $1\sigma$ photon noise and sky estimation (background level) uncertainty. As mentioned above, the residual flat error is negligible and hence is not included here. The latter dominates our error budget.

## Multicomponent decomposition

The traditional method for measuring ICL is to define an SBC and char- acterize the light component fainter than the threshold. As the choice of the threshold is arbitrary, it is difficult to use the method to compare results from different studies. In particular, because the current redshift regime ($1 \lesssim z \lesssim 2$) is considerably different from those of the previous studies, one cannot objectively characterize the ICL properties based on this traditional approach.

In this study, we decompose the BCG + ICL profile into multiple Sérsic components. A Sérsic profile[40] is defined as:

$$I(r) = I_e \exp\left\{-b_n \left[\left(\frac{r}{r_e}\right)^{1/n} - 1\right]\right\},$$

where $r$, $r_e$, $n$ and $I_e$ are the radius, half right radius, Sérsic index and intensity at the half right radius, respectively. In this equation, $b_n$ is the constant that is not independent and is solely determined by $n$.

Then, the BCG + ICL profile $S(r)$ is modelled as a superposition of PSF-convolved multiple Sérsic components $I_m(r)$: $S(r)=\mathrm{PSF}(r)*\sum_m I_m(r)$, where PSF($r$) is the PSF radial profile and the symbol '*' represents the convolution. The PSF radial profile was constructed by combining the core from observed stellar images and the wing from the TinyTim[41] result. If the PSF correction procedure is omitted, the mean ICL fraction increases by around 1.5% (Extended Data Table 2).

How do we determine the total number of Sérsic components for each cluster? A model based on more components has higher degrees of freedom and leads to smaller residuals. However, the drawback is overfitting. About 70% of elliptical galaxies are reported to require more than one Sérsic component to adequately describe their profiles[42–44] In this study, we limit the maximum number of Sérsic components for the description of the BCG + ICL profile to three.

To determine the optimal number of Sérsic components, we use two criteria. The first is the Bayes factor $K$[45], defined as follows:

$$K = \frac{p(D|M_1)}{p(D|M_2)},$$

where $p(D|M_k)$ is the probability of the data ($D$) given the model $M_k$. As $P(D|M_k)$ cannot be computed directly, in practice $K$ is evaluated as follows:

$$K = \frac{p(M_1|D)p(M_2)}{p(M_2|D)p(M_1)} = \frac{p(M_1|D)}{p(M_2|D)},$$

where we assume the equality between the two priors $p(M_1)$ and $p(M_2)$. The Bayes factor $K$ informs us of how the first model $M_1$ is preferred over the second $M_2$. Its outstanding advantage is that $K$ inherently penalizes the model according to its degrees of freedom based on first principles. The second criterion is the number of inflexion points in the derivative of the SB radial profile. Naturally, the existence of $l$ inflexion points implies a preference towards a model with $l + 1$ components. After investigation of our ten clusters with the above two criteria, we find that for the entire sample (1) the Bayes factor $K$ between the best and second-best models is $\ln(K) \gtrsim 2$, (2) the optimal number of Sérsic components inferred from the Bayes factor agrees with the result from the inflexion point analysis and (3) the SB profiles require either two or three components for optimal decomposition.

## Estimation of the total cluster luminosity

One of the key requirements in measuring the ICL fraction is a robust selection of the cluster member galaxies. In this study, the first step towards this goal is the compilation of the results from previous spec- troscopic studies (see Extended Data Table 3). We used this spectroscopic cluster member catalogue to define the initial red sequence locus from the colour–magnitude diagram. By iteratively applying linear regression to the red sequence and selecting objects within 0.5 mag from the best-fit line, we built up the second-stage cluster member catalogue. In this iteration, we removed stars using the CLASS_STAR value reported by SExtractor and the objects either brighter than the BCG or fainter than F105W = 26 mag. This second-stage cluster member catalogue needed to be improved because the distribution of the object distance from the best-fit red sequence line was asymmetric (the blue side was blended with the neighbouring blue cloud). Thus, we fitted a double Gaussian model to the distribution (see the bottom panel of Extended Data Fig. 4). The new centre and $1\sigma$ width of the red sequence were used to update the intercept and width of our previous best-fit linear regression result and the final cluster member catalogue was obtained (see the top panel of Extended Data Fig. 4). Although we employed a sophisticated procedure for selection of cluster members, inevitably the method is designed to select only red members, except for the blue spectroscopic members. If the contribution from the blue members is large, our ICL fraction would be overestimated. However, we argue that the overestimation, if any, would not be significant because (1) in most cases the spectroscopic catalogue includes the brightest blue cluster members (Extended Data Table 3), (2) even the brightest blue cluster members are found to be still a few magnitudes fainter than the BCG and other brightest red members and (3) some fraction of galaxies in our red sequence catalogue are non-cluster members. We verified this claim by utilizing the publicly available photometric redshift catalogue for SPT0205 (ref. 46), which includes the blue cluster member candidates. When we repeated the measurement of the ICL fraction with it, the resulting ICL fraction shifted by only 1%.

The total luminosity is estimated as follows. First, we masked out non-member galaxies/stars from our imaging data. The total masked area is non-negligible and simply assigning zero flux to the area would lead to substantial underestimation of the total luminosity. Thus, we filled the masked regions with the predicted flux from our best-fit multi-Sérsic model. Finally, the total luminosity is computed by the summation of the pixel values of the resulting image, which is comprised of the flux from the BCG, ICL and cluster members.

## Impact of the red sequence selection criteria

Although we took care to robustly define the locus of the red sequence through somewhat sophisticated iteration, two of the remaining ambiguities worth further investigation are the faint-end limit and the width of the red sequence. To examine the dependence of the ICL fraction on these selection criteria, we considered the following three additional cases:

- Test A: the magnitude limit decreased to 24th mag

- Test B: the magnitude limit increased to 28th mag

- Test C: the width increased to $1.5\sigma$

We list the test results for individual clusters in Extended Data Table 3. In the case of Test A, the average (maximum) increase in ICL fraction is found to be approximately 1.9% (approximately 3.8%). Test B shows that the average (maximum) decrease in ICL fraction is approximately 0.9% (approximately 1.6%). Finally, Test C gives an average (maximum) decrease of approximately 2.9% (approximately 5.8%). In summary, our ICL fraction measurements are not sensitive to the selection criteria tested here.

## ICL fraction versus mass correlation

Together with the ICL fraction evolution with redshift, the ICL community has also been investigating the correlation between ICL fraction and halo mass. Under the assumption that massive halos represent older populations, a strong correlation would imply a significant time evolution. Extended Data Fig. 5 shows the ICL fraction versus mass relation for our sample. The masses come from weak lensing studies (Extended Data Table 3). No significant correlation is observed for our sample.

**Online content** Any methods, additional references, Nature Portfolio reporting summaries, source data, extended data, supplementary information, acknowledgements, peer review information, details of author contributions and competing interests, and statements of data and code availability are available at https://doi.org/10.1038/s41586-022-05396-4.

**Data availability** The raw HST near-infrared imaging data used for the current study are publicly available. The processed imaging data are available on the github repository at https://github.com/Hyungjin-Joo/High_z_ICL. Source data are provided with this paper.

**Code availability** An exhaustive repository of code for our custom data processing and analyses reported in this manuscript are available on the github repository at https://github.com/Hyungjin-Joo/High_z_ICL.

**Acknowledgements** This study is based on observations created with NASA/ESA Hubble Space Telescope and downloaded from the Mikulski Archive for Space Telescope at the Space Telescope Science Institute. The current research is supported by the National Research Foundation of Korea under programme 2022R1A2C1003130 and the Yonsei Future-Leading Research Initiative programme.

**Author contributions** M.J.J. conceived, designed and supervised the project. M.J.J. and H.J. analysed the Hubble Space Telescope imaging data, developed the pipeline, interpreted the results and wrote the manuscript.

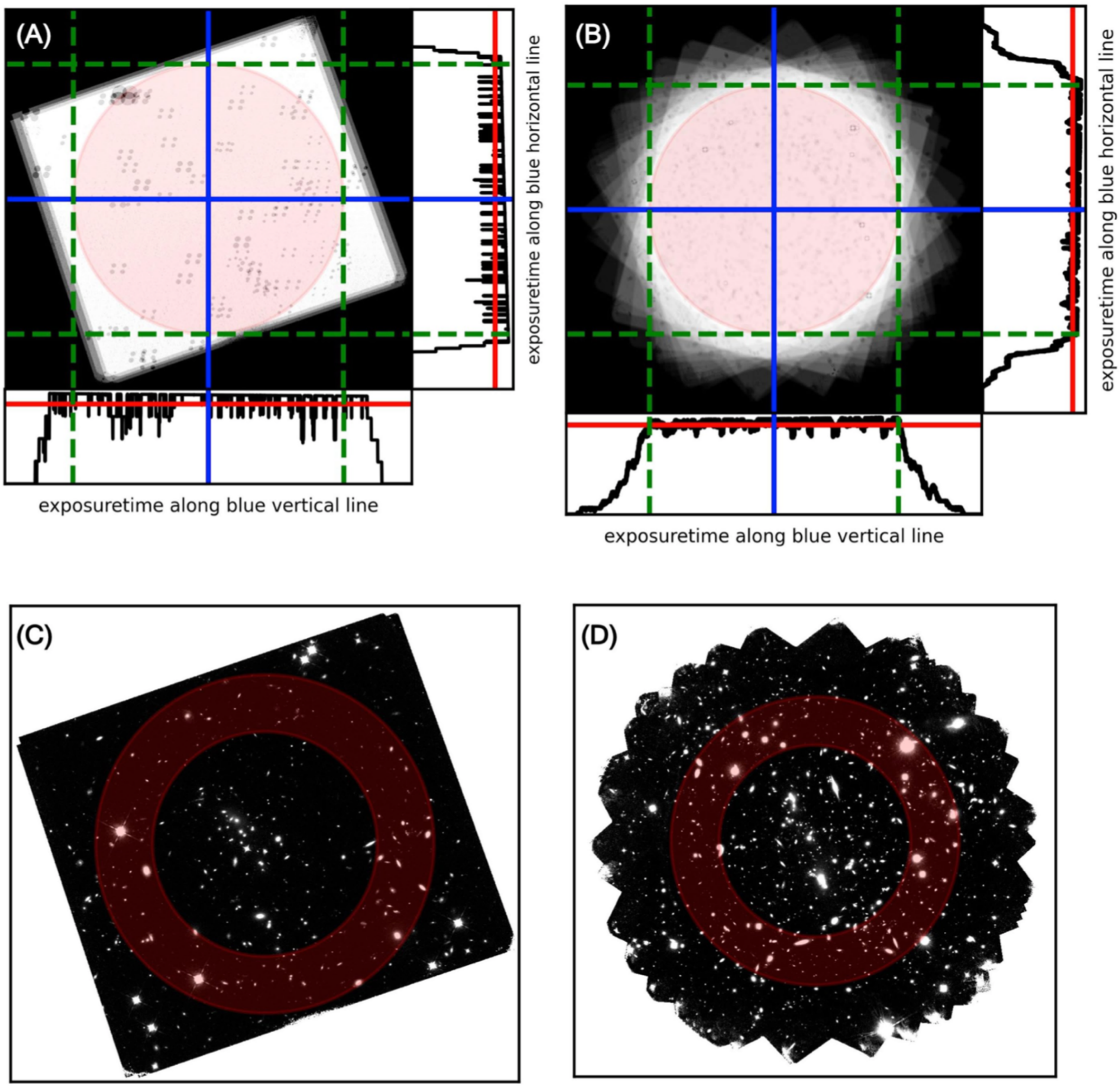

**Extended Data Fig. 1 | Definition of common sky areas.** (A) Exposure map for the single-frame image. (B) Same as (C) except that it is for the mosaic image. (C) Science image for single frame. (D) Same as (C) except that it is for the mosaic image. The pink circular region in (A) is the region that is observed in common by all contributing frames. (B) shows how this common region is positioned in one of the input frames. As the central region of this circle is likely to be heavily influenced by the ICL, we excluded the central region and instead defined the annulus shown in (C) and (D) to estimate the background level.

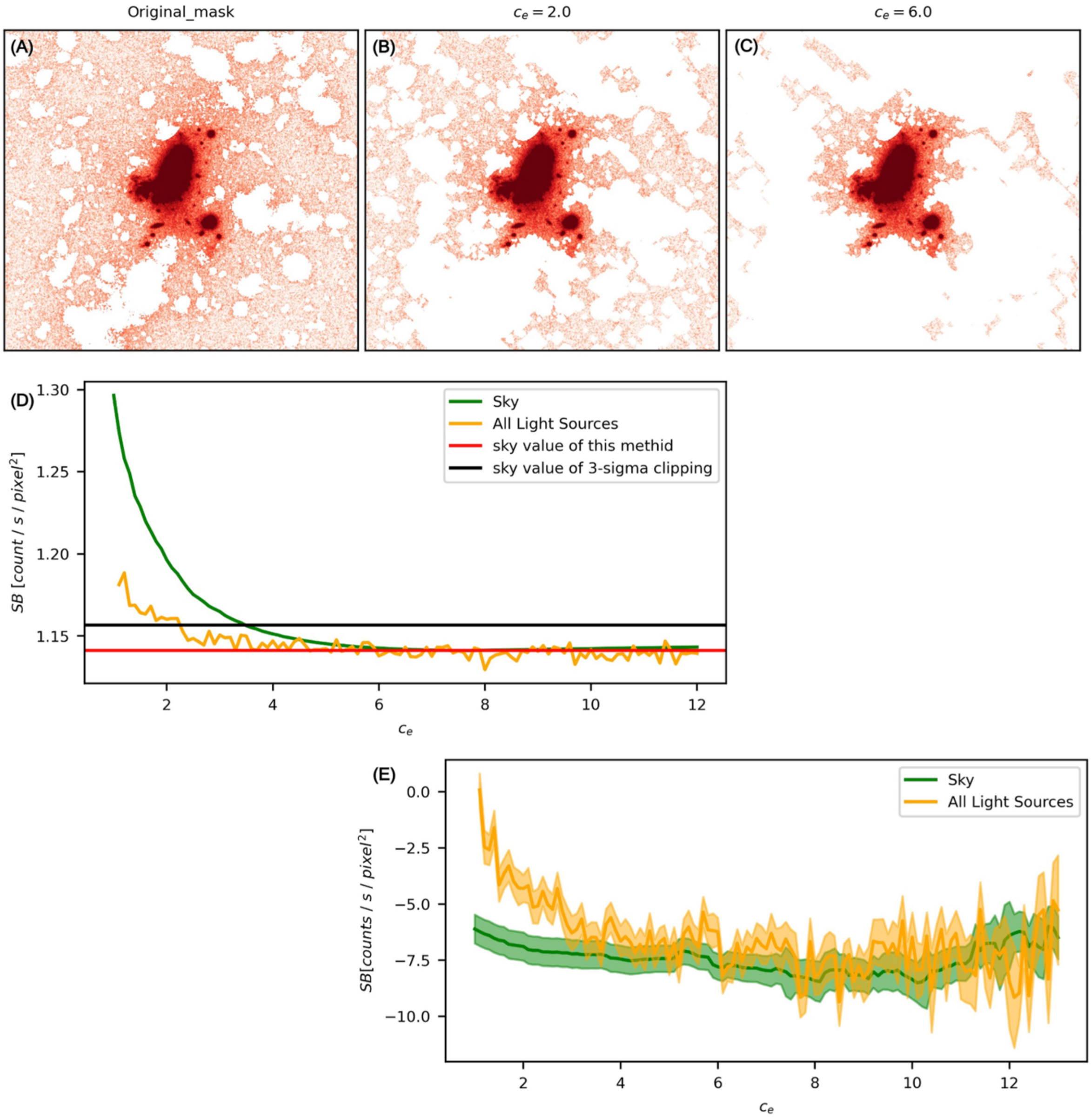

**Extended Data Fig. 2 | Masking size growth and impacts on background level.** (A), (B) and (C) illustrate our scheme for masking size growth from the original to the $c_e = 2$ and $c_e = 6$ cases. Note that we exhaust pixels for ICL measurement at $c_e = 6$. In (D), we show how the background level (green) changes as we vary the masking size using the expansion coefficient for a single exposure. We observe that at $c_e \gtrsim 6$ the measurement converges (red). The black solid line indicates the result when instead we use a $3\sigma$ clipping algorithm without considering the diffuse wings of the astronomical objects. The yellow line shows the surface brightness level measured at each $c_e$. (E) is the same as the left except that the measurement is from the final deep stack. Solid lines indicate the median value and shaded regions show the 68% uncertainty. As the image is deeper, the number of pixels discarded (masked out) at the same $c_e$ value is much greater.

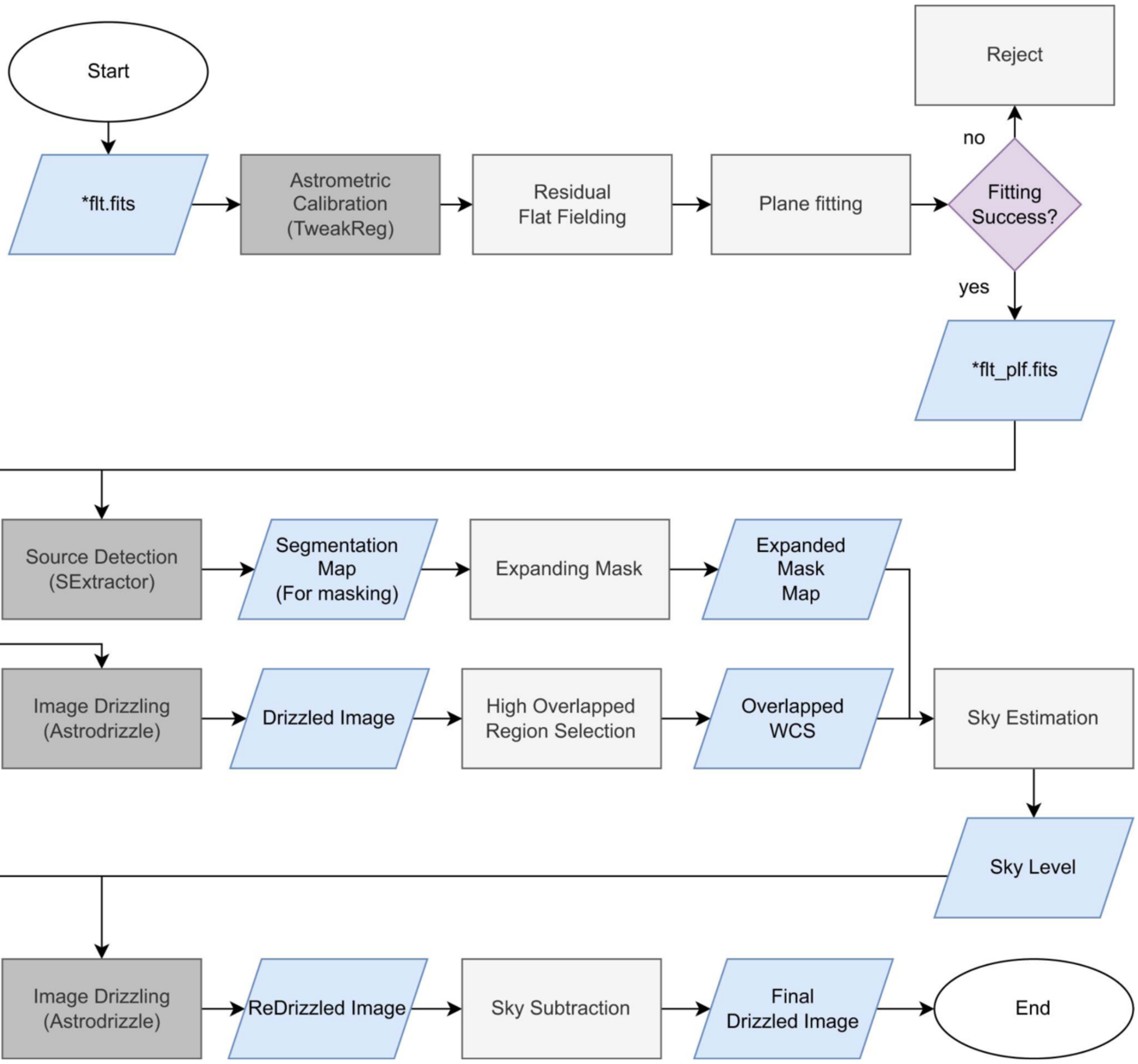

**Extended Data Fig. 3 | Schematic diagram of our ICL-oriented data reduction.** Dark grey rectangles show the steps where external packages are used, while light grey rectangles illustrate our custom procedures. Parallelograms represent the input/output data.

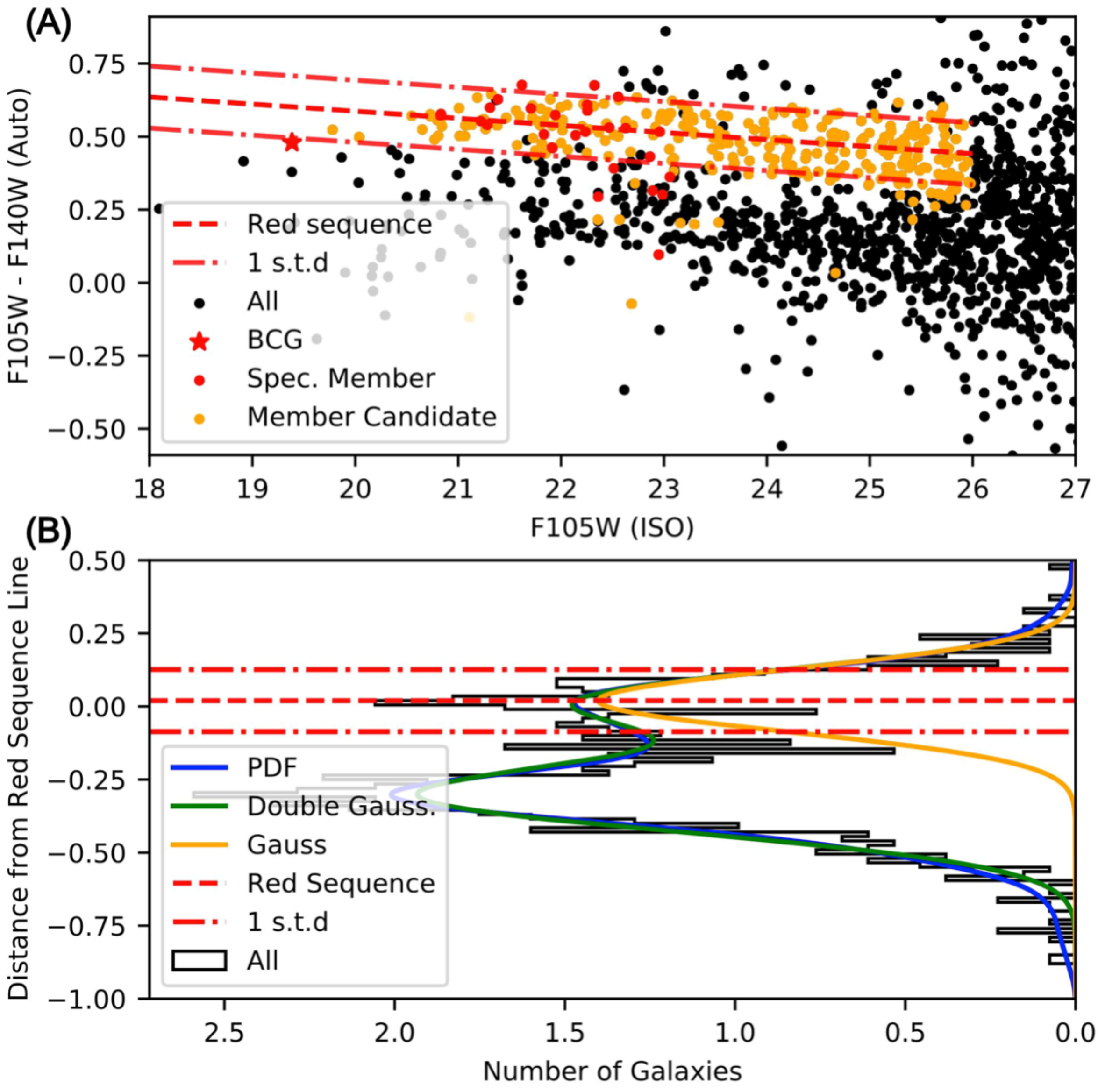

**Extended Data Fig. 4 | Red sequence selection scheme.** Here we display the case for SPT2106. (A) Colour–magnitude diagram. Black dots are all sources detected by SExtractor. The red dots represent the spectroscopic members, whereas the orange dots are our red sequence candidates. The BCG is indicated with a red star. The red dashed line shows the final, best-fit red sequence. The dot-dashed lines bracket the 68% distribution. (B) Distribution of the F105W < 26 object distances from the best-fit red sequence. The green line shows the best-fit double Gaussian models. The yellow line illustrates a single Gaussian component, which represents the distribution of the red sequence candidates.

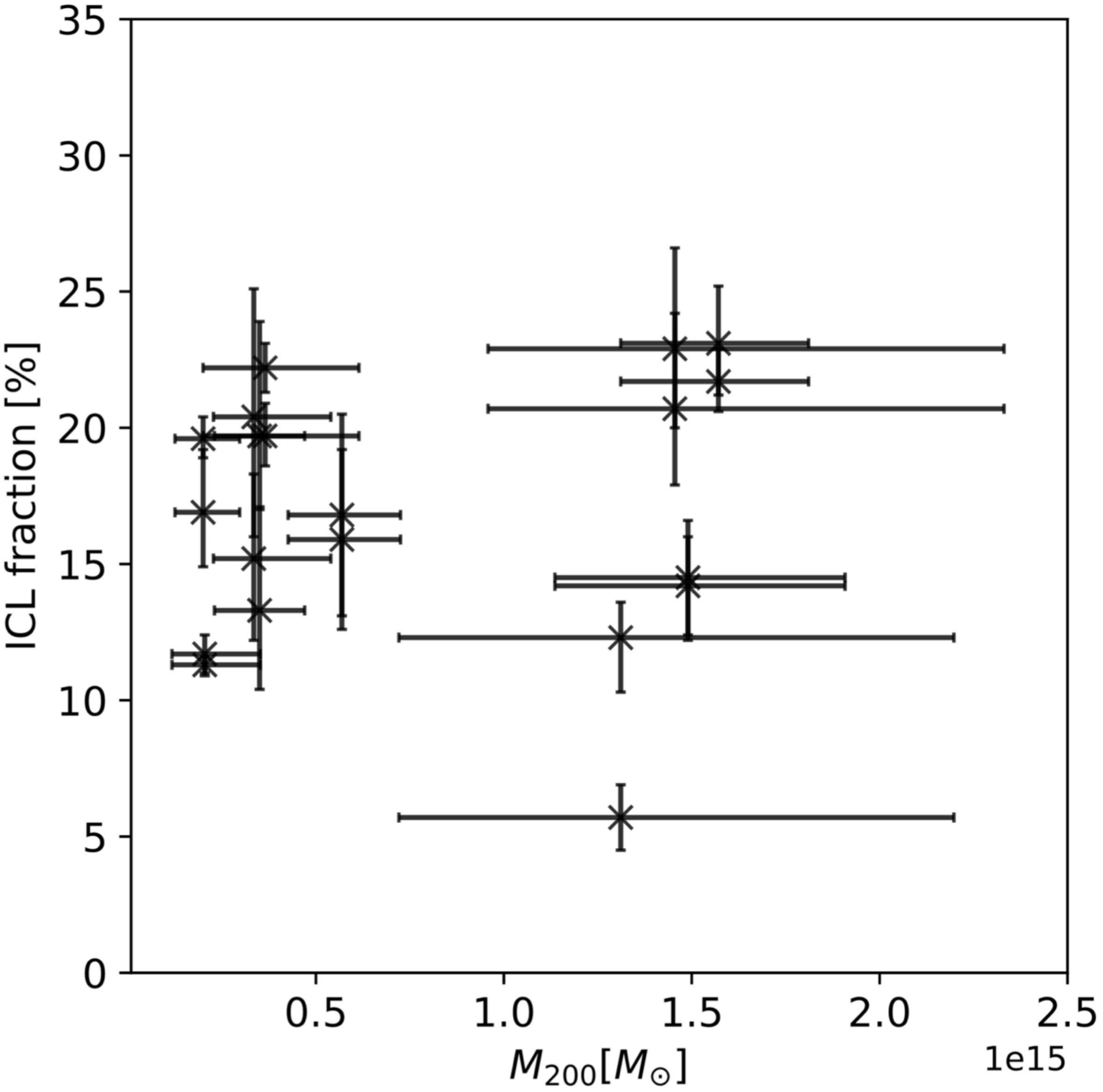

**Extended Data Fig. 5 | Comparison between ICL fraction and cluster mass.** The mass comes from weak lensing studies. No significant correlation between ICL fraction and mass is found.

## Extended Data Table 1 | Target List

| Target (Short Name) | Redshift ($z$) | R.A | Dec | HST Proposal ID | SB limit [mag / arcs$^2$] (Exposure Time [s]) | | |
|---|---|---|---|---|---|---|---|
| | | | | | F150W | F140W | F160W |
| XDCP J1229+0151 (XMM1229) | 0.98 | 12:29:28 | +01:51:34 | 12501 | 29.62 (1,311.7) | - | 29.12 (1,111.7) |
| SPT-CL J2106-5844 (SPT2106) | 1.1312 | 21:06:05 | -58:44:42 | 13677, 14327 | 28.26 (12,567.7) | 28.59 (12,771.6) | - |
| MOO J1142_1529 (MOO1142) | 1.19 | 11:42:46 | +15:27:14 | 14327 | 27.96 (6,283.8) | 28.26 (6,983.8) | - |
| RDCS J1252-2927 (RDCS1252) | 1.237 | 12:52:57 | -29:27:15 | 12501 | 28.26 (1,211.7) | - | 29.00 (1,211.7) |
| MOO J1014+0038 (MOO1014) | 1.24 | 10:14:08 | +00:38:26 | 13677, 14327 | 28.47 (18,255.5) | 28.63 (17,810.1) | - |
| SPT-CL J0205-5829 (SPT0205) | 1.322 | 02:05:46 | -58:29:06 | 13677, 14327 | 28.66 (23,007.2) | 28.84 (25,052.9) | - |
| XDCP J2235-2557 (XMM2235) | 1.39 | 22:35:21 | -25:57:25 | 12501 | 28.18 (1,211.7) | - | 27.85 (1,211.7) |
| SpARCS J1049+5640 (SpARCS1049) | 1.71 | 10:49:22 | +56:40:34 | 13677, 13747 | 28.52 (8,543.3) | - | 28.28 (9,237.4) |
| IDCS J1426.5+3508 (IDCS1426) | 1.75 | 14:26:33 | +35:05:24 | 12203, 13677, 14327 | 30.56 (10,972.4) | 30.79 (11,225.4) | - |
| JKCS041 | 1.803 | 05:26:44 | +04:41:37 | 12927 | 27.73 (2,670.6) | - | 27.99 (4,509.4) |

**Extended Data Table 2 | ICL fractions and impact of various systematics**

| Name | Filter | $f_{BCG+ICL}$ [%] | $f_{ICL}$ [%] | Red Sequence Selection Criteria Test A | Test B | Test C | Unmasked Wings | No PSF |
|---|---|---|---|---|---|---|---|---|
| | | (1) | (2) | (3) | (4) | (5) | (6) | (7) |
| XMM1229 | F105W | $13.6^{+0.5}_{-0.4}$ | $11.7^{+0.7}_{-0.7}$ | 0.8 | -0.3 | -0.8 | -0.2 | 2.5 |
| | F160W | $14.2^{+0.3}_{-0.3}$ | $11.3^{+0.3}_{-0.4}$ | 0.7 | -0.3 | -0.7 | -0.2 | 2.8 |
| SPT2106 | F105W | $18.9^{+1.6}_{-1.6}$ | $14.2^{+1.8}_{-1.8}$ | 0.1 | -0.9 | -2.8 | 0.5 | 0.3 |
| | F140W | $21.4^{+2.2}_{-2.0}$ | $14.5^{+2.1}_{-2.3}$ | 0.1 | -0.8 | -2.7 | 0.3 | 0.6 |
| MOO1142 | F105W | $21.1^{+3.3}_{-3.0}$ | $15.9^{+3.3}_{-3.3}$ | 0.9 | -0.4 | -3.0 | 0.8 | -0.7 |
| | F140W | $24.6^{+3.4}_{-3.3}$ | $16.8^{+3.7}_{-3.7}$ | 1.2 | -0.7 | -4.3 | 0.9 | 3.7 |
| RDCS1252 | F105W | $27.9^{+0.8}_{-0.7}$ | $21.7^{+1.2}_{-1.1}$ | 3.8 | -1.3 | -3.9 | -4.7 | 0.8 |
| | F160W | $29.1^{+1.4}_{-1.2}$ | $24.1^{+2.1}_{-1.9}$ | 2.8 | -0.9 | -3.1 | -3.5 | -0.1 |
| MOO1014 | F105W | $22.5^{+4.7}_{-4.3}$ | $20.4^{+4.7}_{-4.4}$ | 1.4 | -0.5 | -1.9 | 1.1 | -0.5 |
| | F140W | $17.8^{+2.9}_{-2.8}$ | $15.2^{+3.1}_{-3.0}$ | 1.4 | -0.5 | -1.8 | 0.1 | 5.0 |
| SPT0205 | F105W | $17.8^{+2.3}_{-1.9}$ | $16.9^{+2.3}_{-2.0}$ | 1.9 | -0.5 | -3.2 | 0.1 | -3.2 |
| | F160W | $20.7^{+0.9}_{-0.8}$ | $19.6^{+0.8}_{-0.7}$ | 2.0 | -0.5 | -3.3 | 0.2 | -4.2 |
| XMM2235 | F105W | $23.0^{+3.7}_{-2.9}$ | $20.7^{+3.5}_{-2.8}$ | 3.0 | -1.4 | -2.6 | 2.9 | 0.3 |
| | F160W | $25.8^{+2.9}_{-2.9}$ | $22.9^{+3.7}_{-2.9}$ | 3.3 | -1.6 | -2.7 | 2.9 | 2.5 |
| SpARCS1049 | F105W | $27.3^{+4.3}_{-2.7}$ | $19.7^{+4.3}_{-2.7}$ | 2.4 | -1.2 | -5.8 | -1.4 | 4.5 |
| | F160W | $26.6^{+4.0}_{-2.7}$ | $13.3^{+3.8}_{-2.9}$ | 2.5 | -1.1 | -5.2 | -2.0 | 9.7 |
| IDCS1426 | F105W | $22.8^{+1.2}_{-1.1}$ | $19.7^{+1.2}_{-1.1}$ | 2.2 | -0.7 | -3.5 | 0.3 | -1.0 |
| | F140W | $28.4^{+0.8}_{-0.8}$ | $22.2^{+0.9}_{-0.9}$ | 1.9 | -0.7 | -3.5 | -0.8 | 3.2 |
| JKCS041 | F105W | $7.5^{+1.6}_{-0.9}$ | $5.7^{+1.3}_{-1.2}$ | 1.7 | -1.0 | -1.4 | 1.2 | 1.6 |
| | F160W | $16.6^{+0.8}_{-0.9}$ | $12.3^{+1.3}_{-2.0}$ | 3.1 | -1.5 | -2.4 | 1.9 | 0.4 |
| Average | - | 21.38 | 16.89 | 1.86 | -0.89 | -2.93 | 0.02 | 1.46 |

(1) BCG + ICL fraction. (2) Fiducial ICL fraction. (3) Change when the magnitude limit decreased to 24th mag. (4) Change when magnitude limit increased to 28th mag. (5) Change when the width of the red sequence increased to $1.5\sigma$. (6) Change when the expansion coefficient increased to $c_e = 6$. (7) Change when the PSF effect is neglected. For (1) and (2), we quote measurements at $r = 0.5$ Mpc.

**Extended Data Table 3 | Weak lensing mass, the number of spectroscopic member galaxies and their references**

| Name | Weak Lensing Mass ($10^{14} M_{\odot}$) | Number of Spectroscopic Members | Spectroscopic Catalogue Reference |
|---|---|---|---|
| XMM1229 | $2.04^{+1.48}_{-0.87}$ | 17 | [47] |
| SPT2106 | $14.9^{+4.18}_{-3.54}$ | 31 | [46] |
| MOO1142 | $5.69^{+1.56}_{-1.43}$ | 8 | [48] |
| RDCS1252 | $15.7^{+2.40}_{-2.60}$ | 22 | [49] |
| MOO1014 | $3.35^{+2.05}_{-1.07}$ | 7 | [50] |
| SPT0205 | $2.00^{+0.97}_{-0.75}$ | 21 | [46] |
| XMM2235 | $14.6^{+8.76}_{-4.97}$ | 10 | [51] |
| SpARCS1049 | $3.50^{+1.20}_{-1.20}$ | 11 | [52] |
| IDCS1426 | $3.65^{+2.50}_{-1.65}$ | 6 | [53] |
| JKCS041 | $13.1^{+8.87}_{-5.90}$ | 17 | [54] |

## References

1. Rudick, C. S., Mihos, J. C. & McBride, C. K. The quantity of intracluster light: comparing theoretical and observational measurement techniques using simulated clusters. *Astrophys. J.* **732**, 48–64 (2011).
2. Contini, E., De Lucia, G., Villalobos, Á. & Bogani, S. On the formation and physical properties of the intracluster light in hierarchical galaxy formation models. *Mon. Not. R. Astron. Soc.* **437**, 3787–3802 (2014).
3. Burke, C., Collins, C. A., Stott, J. P. & Hilton, M. Measurement of the intracluster at $z \sim 1$. *Mon. Not. R. Astron. Soc.* **425**, 2058–2068 (2012).
4. Ko, J. & Jee, M. J. Evidence for the existence of abundant intracluster light at $z = 1.24$. *Astrophys. J.* **862**, 95–103 (2018).
5. Montes, M. & Trujillo, I. Intracluster light at the Frontier - II. The Frontier Fields Clusters. *Mon. Not. R. Astron. Soc.* **474**, 917–932 (2018).
6. DeMaio, T. et al. The growth of brightest cluster galaxies and intracluster light over the past 10 billion years. *Mon. Not. R. Astron. Soc.* **491**, 3751–3759 (2020).
7. Gonzalez, A. H. et al. Galaxy cluster baryon fractions revisited. *Astrophys. J.* **778**, 14–29 (2013).
8. Presotto, V. et al. Intracluster light properties in the CLASH-VLT cluster MACS J1206.2- 0847. *Astron. Astrophys.* **565**, A126 (2014).
9. Contini, E., Yi, S. K. & Kang, X. Theoretical predictions of colors and metallicity of the intracluster light. *Astrophys. J.* **871**, 24–33 (2019).
10. Burke, C., Hilton, M. & Collins, C. Coevolution of brightest cluster galaxies and intracluster light using CLASH. *Mon. Not. R. Astron. Soc.* **449**, 2353–2367 (2015).
11. Morishita, T. et al. Characterizing intracluster light in the Hubble Frontier Fields. *Astrophys. J.* **846**, 139–151 (2017).
12. Almao-Martinez, K. A. & Blakeslee, J. P. Specific frequencies and luminosity profiles of cluster galaxies and intracluster light in Abell 1689. *Astrophys. J.* **849**, 6–24 (2017).
13. Ellien, A. et al. The complex case of MACS J0717.5+6745 and its extended filament: intra-cluster light, galaxy luminosity function, and galaxy orientations. *Astron. Astrophys.* **628**, A34 (2019).
14. Feldmeier, J. J. et al. Intracluster planetary nebulae in the Virgo Cluster. III. Luminosity of the intracluster light and tests of the spatial distribution. *Astrophys. J.* **615**, 196–208 (2004).
15. Griffiths, A. et al. MUSE spectroscopy and deep observations of a unique compact JWST target, lensing cluster CLIO. *Mon. Not. R. Astron. Soc.* **475**, 2853–2869 (2018).
16. Jee, M. J. Tracing the peculiar dark matter structure in the galaxy cluster Cl 0024+17 with intracluster stars and gas. *Astrophys. J.* **717**, 420–434 (2010).

17. Jimenez-Teja, Y. et al. Unveiling the dynamical state of massive clusters through the ICL fraction. *Astrophys. J.* **857**, 79–96 (2018).

18. Jimenez-Teja, Y. et al. J-PLUS: analysis of the intracluster light in the Coma cluster. *Astrophys. J.* **522**, A183 (2019).

19. Krick, J. E. & Berstein, R. A. Diffuse optical light in galaxy clusters. II. Correlations with cluster properties. *Astrophys. J.* **134**, 466–493 (2007).

20. Mihos, J. C. Intragroup and intracluster light. in *Proc. IAU Symp.: The General Assembly of Galaxy Halos: Structure, Origin and Evolution* vol. 317 (eds Bragaglia, A., Arnaboldi, M., Rejkuba, M. & Romano, D.) 27–34 (Int. Astron. Union, 2015).

21. Yoo, J. et al. Intracluster light properties in a fossil cluster at z = 0.47. *Mon. Not. R. Astron. Soc.* **508**, 2634–2649 (2021).

22. Montes, M. The faint light in groups and clusters of galaxies. *Nature. Astro.* **6**, 308–316 (2022).

23. Navarro, J. F., Frenck, C. S. & White, S. D. M. The structure of cold dark matter halos. *Astrophys. J.* **462**, 563–575 (1996).

24. Asensio, I. A. et al. The intracluster light as a tracer of the total matter density distribution: a view from simulations. *Mon. Not. R. Astron. Soc.* **494**, 1859–1864 (2020).

25. Pillepich, A. et al. First results from the IllustrisTNG simulations: the stellar mass content of groups and clusters of galaxies. *Mon. Not. R. Astron. Soc.* **475**, 648–675 (2018).

26. Murante, G. et al. The diffuse light in simulations of galaxy clusters. *Astrophys. J. Lett.* **607**, 83–86 (2004).

27. Purcell, C. W., Bullock, J. S. & Zentner, A. R. Shredded galaxies as the source of diffuse intrahalo light on varying scales. *Astrophys. J.* **666**, 20–33 (2007).

28. Furnell, K. E. et al. The growth of intracluster light in XCS-HSC galaxy clusters from 0.1 < z < 0.5. *Mon. Not. R. Astron. Soc.* **502**, 2419–2437 (2021).

29. Guennou, L. et al. Intracluster light in clusters of galaxies at redshifts 0.4 < z < 0.8. *Astron. Astrophys.* **537**, A64 (2012).

30. Contini, E. On the origin and evolution of the intra-cluster light: a brief review of the most recent developments. *MDPI.* **9**, 60 (2021).

31. Murante, G. et al. The importance of mergers for the origin of intracluster stars in cosmological simulations of galaxy clusters. *Mon. Not. R. Astron. Soc.* **377**, 2–16 (2007).

32. Contini, E., Yi, S. K. & Kang, E. The different growth pathways of brightest cluster galaxies and intracluster light. *Mon. Not. R. Astron. Soc.* **479**, 932–944 (2018).

33. Tang, L. et al. An investigation of intracluster light evolution using cosmological hydrodynamical simulations. *Astrophys. J.* **859**, 85–97 (2018).

34. DeMaio, T. et al. On the origin of the intracluster light in massive galaxy clusters. *Mon. Not. R. Astron. Soc.* **448**, 1162–1177 (2015).

35. DeMaio, T. et al. Lost but not forgotten: intracluster light in galaxy groups and clusters. *Mon. Not. R. Astron. Soc.* **474**, 3009–3031 (2018).
36. Sahu, K. WFC3 Data Handbook v.5.5 (STScI, 2021).
37. Hoffmann, S. L. et al. The DrizzlePac Handbook v. 2.0 (STScI, 2021).
38. Bertin, E. & Arnouts, S. SExtractor: software for source extraction. *Astron. Astrophys.* **117**, 393–404 (1996).
39. Stone, C. J. et al. AutoProf - I. An automated non-parametric light profile pipeline for modern galaxy surveys. *Mon. Not. R. Astron. Soc.* **508**, 1870–1887 (2021).
40. Sérsic, J. L. Influence of the atmospheric and instrumental dispersion on the brightness distribution in a galaxy. *Boletin de la Asociacion Argentina de Astronomia La Plata Argentina* **6**, 41–43 (1963).
41. Krist, J. Tiny Tim: an HST PSF simulator. *Astronomical Data Analysis Software and Systems II.* **52**, 536 (1993).
42. Oser, L. et al. The two phases of galaxy formation. *Astrophys. J.* **725**, 2312–2323 (2010).
43. Huang, S. et al. The Carnegie-Irvine Galaxy Survey. III. The three-component structure of nearby elliptical galaxies. *Astrophys. J.* **766**, 47 (2013).
44. Huang, S. et al. The Carnegie-Irvine Galaxy Survey. IV. A method to determine the average mass ratio of mergers that built massive elliptical galaxies. *Astrophys. J.* **821**, 114–133 (2016).
45. Gill, J. *Bayesian Methods: A Social Behavioral Science Approch* 2nd edn (CRC, 2008). 46. Balogh, M. L. et al. The GOGREEN and GCLASS surveys: first data release. *Mon. Not. R. Astron. Soc.* **500**, 358–387 (2021).
46. Balogh, M. L. et al. The GOGREEN and GCLASS surveys: first data release. *Mon. Not. R. Astron. Soc.* **500**, 358–387 (2021).
47. Santos, J. S. et al. Multiwavelength observations of a rich galaxy cluster at $z \sim 1$. The HST/ACS colour-magnitude diagram. *Astron. Astrophys.* **501**, 49–60 (2009).
48. Gongalez, A. H. et al. The massive and distant clusters of WISE Survey: MOO J1142+1527, a $10^{15}$ $M_{\odot}$ galaxy cluster at $z = 1.19$. *Astrophys. J. L.* **812**, L40 (2015).
49. Demarco, R. et al. VLT and ACS observations of RDCS J1252.9-2927: dynamical structure and galaxy populations in a massive cluster at $z = 1.237$. *Astrophys. J.* **663**, 164–182 (2007).
50. Decker, B. et al. The massive and distant clusters of WISE Survey. VI. Stellar mass fractions of a sample of high-redshift infrared-selected clusters. *Astrophys. J.* **878**, 72–84 (2019).
51. Santos, J. S. et al. Dust-obscured star formation in the outskirts of XMMU J2235.3-2557, a massive galaxy cluster at $z = 1.4$. *Mon. Not. R. Astron. Soc.* **433**, 1287–1299 (2013).
52. Webb, T. M. A. et al. The star formation history of BCGs to $z = 1.8$ from the SpARCS/SWIRE Survey: evidence for significant in situ star formation at high redshift. *Astrophys. J.* **814**, 96–107 (2015).

53. Stanford, S. A. et al. IDCS J1426.5+3508: discovery of a massive, infrared-selected galaxy cluster at $z = 1.75$. *Astrophys. J.* **753**, 164–171 (2012).
54. Newman, A. B. et al. Spectroscopic confirmation of the rich $z = 1.80$ galaxy cluster JKCS041 using the WFC3 grism: environmental trends in the ages and structure of quiescent galaxies. *Astrophys. J.* **788**, 51–76 (2014).